\documentclass[11pt,a4paper,final]{article}

\usepackage{graphicx}
\usepackage{epstopdf}
\usepackage{amsmath}
\usepackage{amsfonts}
\usepackage{amssymb}
\usepackage[top=2cm,left=3cm,right=2cm,bottom=2cm,includehead,includefoot]{geometry}

\newcommand{\re}{{\mathrm{re}}}
\newcommand{\pre}{{\mathrm{pre}}}

\begin{document}

\title{CMB Constraints on Reheating Models with Varying Equation of State}

\author{Rodolfo~C.~de~Freitas\thanks{rodolfo.camargo (@t) pq.cnpq.br}\\
        \texttt{Instituto Federal de Educa\c{c}\~ao, Ci\^encia e Tecnologia do Esp\'irito Santo,}\\ 
				\texttt{Av. Vit\'oria, 1729, Jucutuquara, Vit\'oria, Brazil}
        \and 
				Sergio~V.~B.~Gon\c{c}alves\thanks{sergio.vitorino (@t) pq.cnpq.br}\\
				\texttt{Universidade Federal do Esp\'irito Santo,}\\ 
				\texttt{Av. Fernando Ferrari, 514, Goiabeiras, Vit\'oria, Brazil}}
				
\date{}

\maketitle

\flushbottom

\begin{abstract}
  The temperature at the end of reheating and the length of this cosmological phase can be bound to the inflationary observables if one considers the cosmological evolution from the time of Hubble crossing until today. There are many examples in the literature where it is made for single-field inflationary models and a constant equation of state during reheating. We adopt two simple varying equation of state parameters during reheating, combine the allowed range of the reheating parameters with the observational limits of the scalar perturbations spectral index and compare the constraints of some inflationary models with the case of a constant equation of state parameter during reheating.
\end{abstract}

\vspace{0.5cm}
\rule{\textwidth}{0.1mm}

\tableofcontents

\vspace{0.5cm}
\rule{\textwidth}{0.1mm}


\newpage

\section{Introduction}
\label{sec:intro}     

The horizon and the flatness problems, the existence of magnetic monopoles and the origin of the large scales structures, \cite{kolb, linde02, muka}, were the great cosmological puzzles of the twentieth century. To solve these problems, an inflationary phase \cite{guth, staro, staro01, sato} has been added to the description of the very early Universe.

Since its establishment, several inflationary testable models were built, providing a description of the Universe in its early stages : new, chaotic, extended, power-law, hybrid, natural, supernatural, extra-natural, eternal, Dterm, F-term, brane, oscillating, trace-anomaly driven, k, ghost, tachyon, etc. One simple and elegant model for inflation was derived and called as chaotic inflation \cite{linde01}, in which the inflaton potential is simply a quadratic function of $\phi$.

The inflationary phase is basically described considering an accelerated expansion of space-time with $\ddot a > 0$, being $a$ the scale factor of the Universe. This condition implies that the comoving Hubble radius $(aH)^{-1}$ decreases with time, while $\dot a = aH $ increases. Considering a homogeneous and isotropic metric, the field equations are given by
\begin{eqnarray}
\label{fieldeq}
(\frac{\dot a}{a})^2 + \frac{K}{a}^2 &=& H^2 + \frac{K}{a}^2 = \frac{8\pi G}{3}\rho\nonumber\quad, \\
\frac{\ddot a}{a} &=& -\frac{4\pi G}{3}(\rho + 3p)\quad,
\end{eqnarray}
where $H$ is the Hubble parameter that indicates the expansion rate of the Universe, $K$ is the spatial curvature, where positive, zero, and negative values correspond respectively to closed, flat, and open spatial sections, $G$ is the Newtonian gravitational constant, $\rho$ is the matter density and $p$ is the pressure. With the previous expression (\ref{fieldeq}), we see that for an accelerated expansion of the Universe, the relation between the density and pressure of the fluid must be $\rho + 3p < 0$, violating the strong energy condition.

To solve a number of problems in the Big Bang model, the inflationary period also brought a number of issues that needed to be answered \cite{linde01, linde03}. Among these issues, the creation of matter and radiation is one of the most challenging. The problem is as follows: The expansion during inflation is so huge that, when inflation ends, any pre-existing matter is very diluted and the Universe is very cold. To return to the situations established by the standard cosmological model, the elements involved in inflation should transmit energy to the relativistic matter that will dominate the next phase of evolution of the Universe. This process of energy exchange at the end of the inflationary era is called reheating and involves changing from a cold and almost matterless Universe to a hot Universe dominated by relativistic particles. This exchange of energy must be such that the present production of entropy within our observable horizon is around $S\approx 10^{89}$ and the matter density is in the order of $\rho\approx 10^{23}~ M_{\odot}$ \cite{lev, bell}.

In most models of inflation, the expansion of the Universe is governed by a classical scalar field $\phi$ called inflaton, which stores the whole energy of the very early Universe \cite{garcia}. At the end of inflation the inflaton begins to oscillate around the minimum of the effective potential and transfers energy to others pre-existing fields through parametric resonance \cite{oli, kof01, gre}, which fills the Universe with particles and radiation. These particles (including photons) begin to interact with each other until the moment when the Universe reaches its thermal equilibrium, which is called thermalization. This is the simplest description of the reheating phase of the Universe. The initial stage of the reheating is called preheating \cite{kof02} and can be divided into two parts: in the first one, when the inflaton interacts with itself and its initial amplitude of oscillations is large enough, the back-reaction process of the created particles can be neglected and the decay occurs exponentially, generating a large occupation numbers in some frequency bands. The time interval for the creation of the particles is generally lower than the expansion of the Universe and we can disregard it. The second stage, where the expansion of the Universe becomes relevant, occurs when the back-reaction and feedback effects becomes significant, being necessary to take into account its interaction with the decaying quantum particles created by harmonic oscillations of the inflaton field in an expanding Universe \cite{kof02}. The preheating stage happens within a scenario where the main mechanism is the broad parametric resonance \cite{dol, bran, bran01} and the end of this phase occurs when the created particles are far away of equilibrium, but with extremely large mean occupation numbers. After the preheating, gradually, the  energy oscillations is transferred into energy of the ultra-relativistic particles. Reheating is completed at the moment when the energy density of the Universe and thermal energy of the created ultra-relativistic particles are equal. The reheating temperature, determined  by  the  total  decay  rate of the inflaton  $\Gamma_{total}$ can be written as $T_r \approx 0.1~\sqrt{\Gamma~M_P}$, where $M_P$ is the Planck mass.

Despite the inflationary phase has several observational constraints, the reheating stage of the Universe is difficult to describe and its theoretical parameters, like the e-folding number $N_{re}$ and the effective equation of state (EoS) parameter $\omega_{re}$, are complicated to relate to the currently available observational data. One way to make this connection is to relate inflation and reheating through the time when the CMB perturbations cross outside the Hubble radius (featuring inflation) and the subsequent time when the perturbations cross inside the Hubble radius (featuring reheating) \cite{Liddle:2003as}. This can be expressed as follows
\begin{equation}
   \label{eq:scale}
   \ln{\left(\frac{k}{a_0 H_0}\right)} = -N_k - N_\re +\ln{\left(\frac{a_{\mathrm{eq}} H_{\mathrm{eq}}}{a_0 H_0}\right)} + \ln{\left(\frac{H_k}{H_{\mathrm{eq}}}\right)} \quad,
\end{equation}
where $k$ is the wavenumber, $N_k$ is the number of e-folds between horizon crossing and the end of inflation, $N_\re$ and $N_{\mathrm{RD}}$ are the number of e-folds between the end o inflation and the end o reheating and between the end of reheating and the end of radiation-dominated era, respectively. Thus it is clear that we can use the observable parameters of CMB to obtain constrains associated with the reheating era. This technique has been broadly used in the literature \cite{Martin:2006rs,Lorenz:2007ze,Martin:2010kz,Adshead:2010mc,Mielczarek:2010ag,Easther:2011yq,Dai:2014jja,Martin:2014nya,Cook:2015vqa,Munoz:2014eqa,Cai:2015soa,Gong:2015qha} assuming a constant equation of state parameter during reheating. In this work we use the same technique in order to calculate the length of reheating and the temperature at the end of the era in some single-field inflationary models, adopting a step-like function and a continuous time dependent equation of state during reheating.

This work is organized as follows: In section \ref{sec:reheating} we present three models for the equation of state parameter during reheating and use them to relate the duration of the reheating phase and the reheating temperature to the inflationary parameters. In section \ref{sec:inflation}, we review the polynomial, Starobinsky and Higgs inflationary models and see how they constraint reheating. And finally, in section \ref{sec:conclusions}, we present our conclusions.

\section{Reheating parameters}
\label{sec:reheating}         

The comoving Hubble scale $a_k H_k = k$ when the primordial fluctuations crossed the Hubble radius during inflation can be related to the present scale \cite{Liddle:2003as},
\begin{equation}
   \frac{k}{a_0 H_0} = \frac{a_k}{a_{\mathrm{end}}} \frac{a_{\mathrm{end}}}{a_\re} \frac{a_\re}{a_{\mathrm{eq}}} \frac{a_{\mathrm{eq}} H_{\mathrm{eq}}}{a_0 H_0}\frac{H_k}{H_{\mathrm{eq}}} \, .
\end{equation}

Here, $k$ is the scale of horizon exit and can be fixed as the pivot scaled at which a specific CMB experiment determines the scalar spectral index $n_S$. We will use Planck's pivot scale of $k=0.05~\mathrm{Mpc}^{-1}$ \cite{Ade:2015xua,Ade:2015lrj}. ``$\mathrm{end}$'', ``$\re$'', ``$\mathrm{eq}$'' and ``$0$'' mean the end of inflation, the end of reheating, the time of radiation-matter equality and today, respectively. Defining $N_k = \ln{(a_{\mathrm{end}}}/a_k)$ as the number of e-folds between horizon crossing and the end of inflation, $N_\re = \ln{(a_\re/a_{\mathrm{end}})}$ and $N_{\mathrm{RD}} = \ln{(a_{\mathrm{eq}}/a_\re)}$ as the number of e-folds between the end of inflation and the end of reheating and between the end of reheating and the end of radiation-dominated era, respectively, we can find the relation
\begin{equation}
      \ln{\left(\frac{k}{a_0 H_0}\right)} = -N_k - N_\re +\ln{\left(\frac{a_{\mathrm{eq}} H_{\mathrm{eq}}}{a_0 H_0}\right)} + \ln{\left(\frac{H_k}{H_{\mathrm{eq}}}\right)} \,.
\end{equation}

For a given inflationary model and from observational constraints on the primordial power spectrum we can calculate $N_k$. With help of equation (\ref{eq:scale}) and another expression that gives the post-inflationary evolution of the energy density and temperature we can calculate $N_\re$ and $T_\re$.

The energy density at the end of reheating is given by
\begin{equation}
   \label{eq:density_reheating}
	  \rho_\re = \frac{\pi^2}{30}g_\re T_\re^4 \,,
\end{equation}
where $g_\re$ is the number of relativistic species at the end of reheating, which we will assume $g_\re = 100$. The subsequent density evolution is mainly driven by radiation until the radiation-matter equality time. Assuming that the entropy is conserved between the end of reheating and the present time we can relate the reheating temperature to the temperature today with the relation
\begin{equation}
   \label{eq:T1}
   T_\re = T_0 \left(\frac{a_0}{a_\re}\right)\left(\frac{43}{11g_\re}\right)^{1/3} \, .
\end{equation}

In equation (\ref{eq:T1}) the ratio $a_0/a_\re$ can be rewritten as
\begin{equation}
   \frac{a_0}{a_\re} = \frac{a_0}{a_{\mathrm{eq}}}\frac{a_\mathrm{eq}}{a_\re} = \frac{a_0}{a_{\mathrm{eq}}} e^{N_{\mathrm{RD}}} \,,
\end{equation}
and assuming the horizon crossing condition $k=a_k H_k$, the ratio $a_0/a_{\mathrm{eq}}$ can be expanded as
\begin{equation}
   \frac{a_0}{a_{\mathrm{eq}}} = \frac{a_0 H_k}{k}e^{-N_k}e^{-N_\re}e^{-N_{\mathrm{RD}}} \,,
\end{equation}
and we can finally write an expression for the reheating temperature $T_\re$ that is a function of the inflationary model and the cosmic evolution between the end of inflation and the beginning of radiation-dominated era, i. e., during reheating,
\begin{equation}
   \label{eq:T_reheating}
   T_\re = \left(\frac{43}{11g_\re}\right)^{1/3}\left(\frac{a_0T_0}{k}\right)H_k e^{-N_k}e^{-N_\re} \,.
\end{equation}
Notice, from equation (\ref{eq:T_reheating}), that the shorter the reheating era, the larger will be the temperature at the end of reheating.

The energy density evolution during reheating is determined with help of the conservation relation $\dot{\rho}+3H(1+\omega)\rho=0$, where over-dot means time derivative and $\omega = p/\rho$ is the equation of state parameter. Integrating from the end of inflation (beginning of reheating) until the end of reheating one finds
\begin{equation}
   \label{eq:continuity}
	 \ln{\frac{\rho_\re}{\rho_{\mathrm{end}}}} = -3 \int_{a_{\mathrm{end}}}^{a_\re}{(1+\omega)\frac{da}{a}} \, .
\end{equation}

In the next subsections we will calculate the energy density evolution during reheating for three different, but equivalent, models.

\subsection{Constant equation of state parameter}
\label{subsec:model1}                            

Assuming a constant equation of state parameter of value $\omega_\re$ one can easily find that
\begin{equation}
   \label{eq:density1}
   \ln{\left(\frac{\rho_\re}{\rho_{\mathrm{end}}}\right)} = 3(1+\omega_\re)\ln{\left(\frac{a_\re}{a_{\mathrm{end}}}\right)} \,.
\end{equation}
In section \ref{sec:inflation} we will show that the energy density at the end of inflation is $\rho_{\mathrm{end}}=2V_{\mathrm{end}}/(1-\omega_{\mathrm{end}})$. Plugging the temperature equation (\ref{eq:T_reheating}) into the density equation (\ref{eq:density_reheating}) and substituting the result in equation (\ref{eq:density1}) one can find, for $\omega_\re \neq 1/3$
\begin{equation}
   \label{eq:Nre1}
   N_\re = \frac{4}{1-3\omega_\re}\left[-\frac{1}{4}\ln{\left(\frac{45}{\pi^2 g_\re}\right)}-\frac{1}{3}\ln{\left(\frac{11g_\re}{43}\right)}-\ln{\left(\frac{k}{a_0T_0}\right)}-\ln{\left(\frac{V_{\mathrm{end}}^{1/4}}{H_k}\right)}-N_k\right] \,,
\end{equation}
where we used $N_\re = \ln{(a_\re/a_{\mathrm{end}})}$ and assumed $\omega_{\mathrm{end}} = -1/3$. Notice that, due to the quantities $H_k$, $N_k$ and $V_{\mathrm{end}}$, equation (\ref{eq:Nre1}) depends on the chosen inflationary model and is independent of $T_\re$. Plugging equation (\ref{eq:Nre1}) back into the temperature equation (\ref{eq:T_reheating}) one can find a $T_\re$ that depends on the inflationary model and $\omega_\re$.

This result is already known in the literature \cite{Dai:2014jja,Martin:2014nya,Cook:2015vqa,Munoz:2014eqa,Cai:2015soa,Gong:2015qha} and is reproduced here for comparison.

\subsection{Step function equation of state parameter}
\label{subsec:model2}                                 

Inflation ends when potential and kinetic energy are comparable, which makes the equation of state parameter raise from $-1$ to $-1/3$. At this point the equation of state parameter starts to oscillate between $-1$ (potential dominance) and $1$ (kinetic dominance), while it decays toward $0$, which is  the equation of state parameter value at the beginning of reheating. As the inflaton transfers its energy to some other field, the equation of state parameter evolves from $0$ to $1/3$, when begins radiation dominance. It was shown that in a simple $\phi^2$ inflationary model, which interacts with the decay product $\chi$ via $g^2\phi^2 \chi^2$, the equation of state parameter evolves abruptly from $0$ to $0.2$ -- $0.3$ \cite{Podolsky:2005bw}. Although the equation of state parameter jumps from $0$ to radiation domination almost instantaneously, which allow us to treat $\omega_\re$ as a constant during the whole reheating process, we will take into consideration this abrupt transition. The price to pay are new parameters that give shape to the jump.

Assuming a reheating model where the equation of state parameter changes discontinuously, we write
\begin{equation}
   \label{eq:EoS2}
   \omega = 
	    \begin{cases}
			    \omega_\pre  \,, \quad a_{\mathrm{end}} \leq a \leq a_\pre \\
				  \omega_\re \,, \quad a_\pre \leq a \leq a_\re
			\end{cases}
			\, .
\end{equation}
Following \cite{Podolsky:2005bw}, we will adopt $\omega_\pre = 0$ and $\omega_\re = 0.25 \mbox{ -- } 0.33$. With the definitions $N_\re = \ln{(a_\re/a_{\mathrm{end}})}$ and $N_\pre = \ln{(a_\pre/a_{\mathrm{end}})}$ and solving the conservation equation (\ref{eq:continuity}) with the equation of state parameter (\ref{eq:EoS2}) we find
\begin{equation}
   \label{eq:Nre2_1}
   N_\re = \frac{1}{3(1+\omega_\re)}\left[\ln{\left(\frac{\rho_{\mathrm{end}}}{\rho_\re}\right)}+3(\omega_\re-\omega_\pre)N_\pre\right] \,.
\end{equation}
Note that $N_\re = N_\pre + \ln{(a_\re/a_\pre)}$, and equation (\ref{eq:Nre1}) can be recovered if we make $\omega_\pre = \omega_\re$. When compared with the constant equation of state parameter model, subsection \ref{subsec:model1}, we have here the new parameter $N_\pre$, which tell us how long the reheating stage will be in the $\omega_\pre=0$ regime.

Plugging again the temperature equation (\ref{eq:T_reheating}) into the density equation (\ref{eq:density_reheating}) and substituting the result in equation (\ref{eq:Nre2_1}) we find, for $\omega_\re \neq 1/3$
\begin{eqnarray}
   \label{eq:Nre2}
   & N_\re & = \frac{4}{1-3\omega_\re}\bigg{[}-\frac{1}{4}\ln{\left(\frac{45}{\pi^2 g_\re}\right)}-\frac{1}{3}\ln{\left(\frac{11g_\re}{43}\right)}-\ln{\left(\frac{k}{a_0T_0}\right)}-\ln{\left(\frac{V_{\mathrm{end}}^{1/4}}{H_k}\right)} \nonumber \\ 
	&& -N_k + \frac{3}{4}(\omega_\pre - \omega_\re)N_\pre\bigg{]} \,,
\end{eqnarray}
where we assumed again $\omega_{\mathrm{end}} = -1/3$. Notice, another time, that equation (\ref{eq:Nre2}) depends on the chosen inflationary model and is independent of $T_\re$, which can be found plugging equation (\ref{eq:Nre2}) back into the temperature equation (\ref{eq:T_reheating}).

\begin{figure}[!ht]
   \begin{center}
      \includegraphics[width=0.85\textwidth]{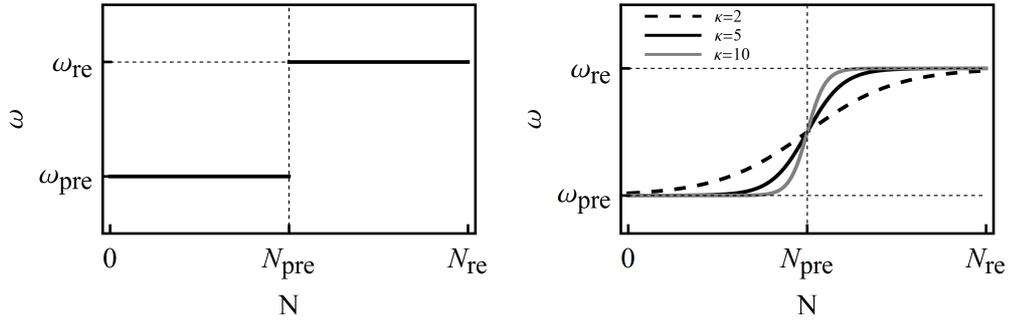}
      \caption{The equation of state parameter. In the left panel, the model described by equation (\ref{eq:EoS2}) and in the right panel, the model described by equation (\ref{eq:EoS3}).}     
      \label{fig:EoS}
   \end{center}
\end{figure}

\subsection{Time dependent equation of state parameter}
\label{subsec:model3}                                  

In this third model we will adopt an equation of state parameter that, as in subsection \ref{subsec:model2}, changes abruptly, but this time in a continuous manner. We choose to model the equation of state parameter as

\begin{equation}
   \label{eq:EoS3}
	 \omega = \frac{\omega_\re}{2}\left\{1+\tanh{\left[\kappa(N_\re-N_\pre)\right]}\right\} \,,
\end{equation}
which means that the equation of state parameter changes from $\omega = 0$ to $\omega = \omega_\re$ around $N_\pre$, and $\kappa$ is the model parameter that controls how sharp this change is made. In figure \ref{fig:EoS} we plot $\omega$ for equations (\ref{eq:EoS2}) and (\ref{eq:EoS3}).

With the equation (\ref{eq:EoS3}) into (\ref{eq:continuity}) we find
\begin{equation}
   \label{eq:Nre3}
   e^{3(1+\omega_\re)N_\re}\left[1+e^{-2\kappa(N_\re-N_\pre)}\right]^{\frac{3\omega_\re}{2\kappa}}=\frac{\rho_{\mathrm{end}}}{\rho_\re}\left[1+e^{2\kappa N_\pre}\right]^{\frac{3\omega_\re}{2\kappa}} \, ,
\end{equation}
where the ratio between the energy densities is
\begin{equation}
   \label{eq:ratio_densities}
   \frac{\rho_{\mathrm{end}}}{\rho_\re} = \left(\frac{30}{\pi^2 g_\re}\right)\left(\frac{11 g_\re}{43}\right)^{4/3}\left(\frac{k}{a_0 T_0}\right)\left(\frac{2}{1-\omega_{\mathrm{end}}}\right)\left(\frac{V_{\mathrm{end}}}{H_k^4}\right)e^{4N_k}e^{4N_\re} \,.
\end{equation}
Equation (\ref{eq:Nre3}) will be numerically solved for $N_\re$, with $N_\pre$, $\kappa$ and $\omega_\re$ fixed. 

\begin{figure}[!ht]
   \begin{center}
      \includegraphics[width=0.85\textwidth]{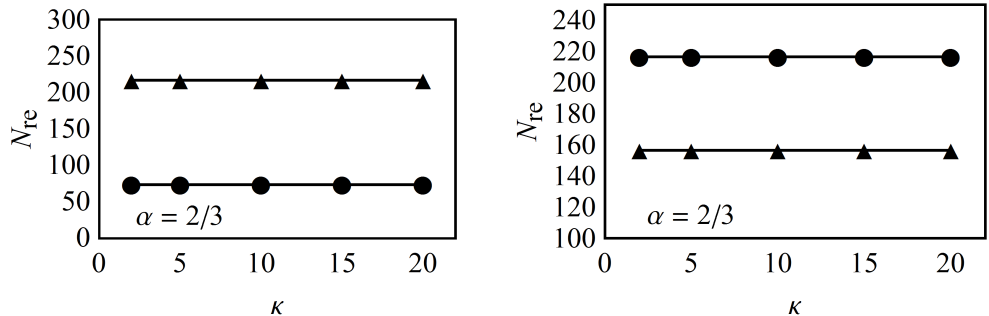}
      \caption{The length of reheating $N_{\re}$ for the time dependent equation of state parameter model, polynomial inflationary potential of index $\alpha = 3/2$ and $n_S=0.9682$. In the left panel we fixed $N_\pre=25$, the dots and the triangles are for $\omega_\re = 0$ and $\omega_\re=0.25$, respectively. In the right panel we fixed $\omega_\re=0.25$, the dots and triangle are for $N_\pre=25$ and $N_\pre=45$, respectively.}     
      \label{fig:Poly_3_kappa}
   \end{center}
\end{figure}

Looking at equation (\ref{eq:Nre3}) we can presume that it do not varies with $\kappa$ or, at least, depends weakly on it. In figure \ref{fig:Poly_3_kappa} we show how $N_\re$ varies with $\kappa$, in the interval $(2,20)$, for two different values of $N_\pre$ and $\omega_\re$, with a polynomial inflationary potential of index $\alpha = 2/3$, where we adopt the Planck central value for the primordial scalar spectral index $n_S$ \cite{Ade:2015lrj}. More details about the inflationary models dependence will be given in next section. As the reader can guess by looking at equation (\ref{eq:Nre3}), the behavior showed in figure \ref{fig:Poly_3_kappa} is independent of the chosen inflationary potential, therefore, we will not repeat here the plots for other potentials. As a matter of computational time, we will adopt $\kappa = 2$ in our analysis.

Although equation (\ref{eq:Nre3}) seems to be quite different when compared with equation (\ref{eq:Nre2_1}), it is not difficult to show that they are equivalent. We can rewrite the expression (\ref{eq:Nre3}) as

\begin{equation}
   \label{eq:Nre3_1}
   3(1+\omega_\re)N_\re +\frac{3\omega_\re}{\kappa}\ln{\left[\frac{1+e^{-2\kappa(N_\re-N_\pre)}}{1+e^{2\kappa N_\pre}}\right]} = \ln{\left(\frac{\rho_{\mathrm{end}}}{\rho_\re}\right)} \,.
\end{equation}

Taking into account that $N_\re$ and $N_\pre$ are typically of order unity or bigger we can show that in the limit of big $\kappa$ the second term at the left hand side of equation (\ref{eq:Nre3_1}) will approximate

\begin{equation}
   \label{eq:kappa_infity}
   \lim_{\kappa \rightarrow \infty}{ \frac{1}{\kappa}\ln{\left[\frac{1+e^{-2\kappa(N_\re-N_\pre)}}{1+e^{2\kappa N_\pre}}\right]}} \rightarrow - N_\pre \, ,
\end{equation}
and finally we have
\begin{equation}
   \label{eq:Nre3_approx}
   3(1+\omega_\re)N_\re -3\omega_\re N_\pre \approx \ln{\left(\frac{\rho_{\mathrm{end}}}{\rho_\re}\right)} \,,
\end{equation}
which is the same as equation (\ref{eq:Nre2_1}), with $\omega_\pre = 0$. Even so, equations (\ref{eq:Nre2}) and (\ref{eq:Nre3}) can drive us into different results as we will see in the next section.

\section{Inflationary models}
\label{sec:inflation}        

Once we have established the models for the cosmic evolution during reheating, we can now review the single field slow-roll inflation, at first for a general inflaton potential $V(\phi)$. For more details about inflation, the reader can refer to the reviews \cite{Riotto:2002yw,Baumann:2009ds}. After that, we can calculate $N_k$, $H_k$ and $V_{\mathrm{end}}$ for the specific inflationary models.

A Universe filled with a single scalar field $\phi$ is described by the equations
\begin{eqnarray}
   && 3H^2=M_{\mathrm{Pl}}^{-2}\left(\frac{\dot{\phi}^2}{2}+V(\phi)\right) \,, \\
	 && \ddot{\phi}+3H\dot{\phi}+V^{\prime} = 0 \,,
\end{eqnarray}
where over-dot is the derivative of cosmic time, prime is the derivative of $\phi$, $H$ is the Hubble parameter and $M_{\mathrm{Pl}}$ is the Planck mass. 

The slow-roll parameters are
\begin{eqnarray}
   \label{eq:epsilon}
   && \epsilon \equiv -\frac{\dot{H}}{H^2} = \frac{M^{-2}_{\mathrm{Pl}}}{2}\frac{\dot{\phi}^2}{H^2} \approx \frac{M_{\mathrm{Pl}}^2}{2}\left(\frac{V^\prime}{V}\right)^2 \ll 1 \,, \\
	 \label{eq:eta}
	 && \eta \equiv -\frac{\ddot{\phi}}{H\dot{\phi}} \approx M_{\mathrm{Pl}}^2\frac{V^{\prime \prime}}{V} \ll 1 \,.
\end{eqnarray}
The smallness of $\epsilon$ and $\eta$ guarantees that the inflaton potential dominates over the the kinetic energy during inflation, and that cosmic expansion is accelerated and quasi-exponential. Conditions (\ref{eq:epsilon}) and (\ref{eq:eta}) mean that
\begin{eqnarray}
   && 3H^2 \approx M_{\mathrm{Pl}}^{-2}V \,, \\
	 && 3H\dot{\phi}+V^\prime \approx 0 \, ,
\end{eqnarray}
and the number of $e$-folds between horizon crossing and the end of inflation is
\begin{equation}
   \label{eq:Nk}
	 N_k = \int_{t_k}^{t_{\mathrm{end}}}{Hdt} \approx \int_{\phi_{\mathrm{end}}}^{\phi_k}{\frac{V}{V^\prime}d\phi} \,.
\end{equation}

The primordial scalar perturbations power spectrum can be written as
\begin{equation}
   P_S = A_S\left(\frac{k}{k_{\mathrm{p}}}\right)^{n_S-1} \,,
\end{equation}
where $A_S$ is the Planck scalar power spectrum amplitude with central value $A_S = 2.196\times 10^{-9}$, $k_{\mathrm{p}} = 0.05~\mathrm{Mpc}^{-1}$ is Planck's pivot scale \cite{Ade:2015lrj} and the scalar spectral index $n_S$ is
\begin{equation}
   \label{eq:nS}
	 n_s-1 = 6\epsilon - 2\eta \,.
\end{equation}

The ratio between the power spectrum of the primordial tensor perturbations and that of the scalar perturbations at horizon crossing is
\begin{equation}
   r_k = \frac{P_t}{P_S} = \frac{2H_k^2}{\pi^2M_{\mathrm{Pl}}^2A_S} \,.
\end{equation}
Using the slow-roll relation $r_k \approx 16\epsilon_k$ one can find
\begin{equation}
   \label{eq:Hk}
   H_k \approx \pi M_{\mathrm{Pl}}\sqrt{8A_S\epsilon_k} \,.
\end{equation}

Inflation ends when $\epsilon \approx 1$, which is equivalent to $\omega_{\mathrm{end}} = -1/3$ and means that at the end of inflation the inflaton's kinetic energy is comparable to that of the potential energy. This allow us to find
\begin{equation}
   \rho_{\mathrm{end}} = \left(\frac{2}{1-\omega_{\mathrm{end}}}\right)V_{\mathrm{end}} \,.
\end{equation}

The form of the potential $V_{\mathrm{end}}$ can be determined as a function of the model parameters at pivot scale once the inflationary potential $V(\phi)$ is chosen. 

\subsection{Polynomial potentials}
\label{subsec:polynomial}         

We now apply the calculations developed above to a single scalar field of polynomial potential of index $\alpha$
\begin{equation}
   V(\phi)=\frac{m^{4-\alpha}}{2}\phi^\alpha \,,
\end{equation}
where $m$ is the inflaton field mass. Applying equation (\ref{eq:Nk}) we find the number of $e$-folds between horizon crossing and the end of inflation
\begin{equation}
   \label{eq:Nk_Poly1}
	  N_k = \frac{1}{2\alpha M_{\mathrm{Pl}}}\left(\phi_k^2-\phi_{\mathrm{end}}^2\right) \approx \frac{\phi_k^2}{2\alpha M_{\mathrm{Pl}}} \,,
\end{equation}
where in the last step we take in account that in these models $\phi_k \gg \phi_{\mathrm{end}} $. We can now use equations (\ref{eq:epsilon}) and (\ref{eq:eta}) to find the slow-roll parameters $\epsilon$ and $\eta$ as a function of $\phi_k$. Using equation (\ref{eq:nS}) for the spectral index and substituting the result at equation (\ref{eq:Nk_Poly1}), we finally find
\begin{equation}
   N_k \approx \frac{\alpha+2}{2(1-n_S)} \,.
\end{equation}

With the same procedure we can use equation (\ref{eq:Hk}) to find
\begin{equation}
   \label{eq:Hk_Poly}
	 H_k \approx \pi M_{\mathrm{Pl}}\sqrt{\frac{4\pi A_S}{\alpha+2}(1-n_S)} \,.
\end{equation}

The scalar field potential at the end of inflation can be found with help of equation (\ref{eq:Hk_Poly}) and taking into account that $\epsilon_{\mathrm{end}}=1$, which leads to
\begin{equation}
   \label{eq:Vend_Poly}
	 V_{\mathrm{end}} = \frac{m^{4-\alpha}}{2}\phi_k^\alpha\frac{\phi_{\mathrm{end}}^\alpha}{\phi_k^{\alpha}} \approx 3\pi^2M_{\mathrm{Pl}}^4A_S\frac{\alpha(1-n_S)^2}{\alpha+2} \,.
\end{equation}

\begin{figure}[!ht]
   \begin{center}
      \includegraphics[width=0.85\textwidth]{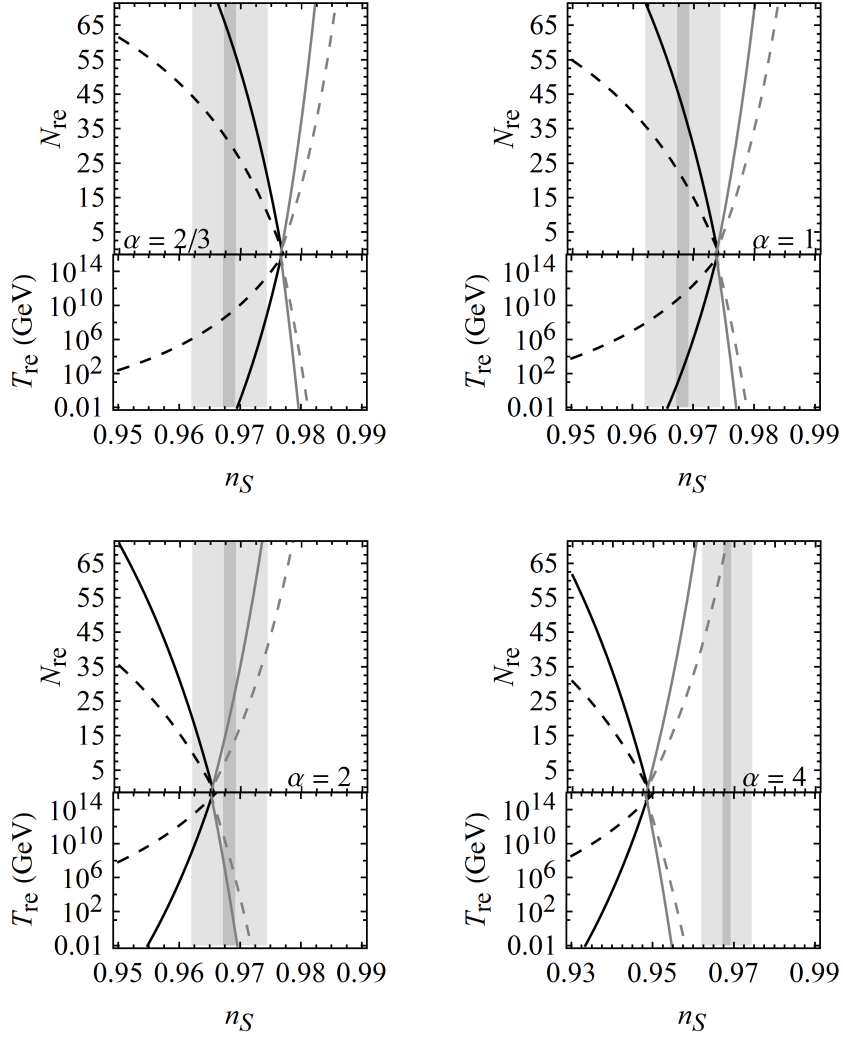}
		  \caption{The length of reheating $N_{\re}$ and temperature at the end of reheating $T_\re$, for the model with constant equation of state parameter and polynomial inflationary potential of index $\alpha$. The black dashed, black solid, gray dashed and gray solid curves correspond to $\omega_\re$ equals to $-1/3$, $0$, $2/3$ and $1$, respectively. The light gray shaded region represents the $1\sigma$ bounds on the primordial power spectrum index $n_{\mathrm S}$ from Planck, while the dark gray shaded region correspond to the $1\sigma$ bound of a CMB experiment with sensitivity $\pm 10^{-3}$ and the same central value for $n_{\mathrm S}$ as Planck.}
      \label{fig:Poly_1}
   \end{center}
\end{figure}

\begin{figure}[!ht]
   \begin{center}
      \includegraphics[width=0.85\textwidth]{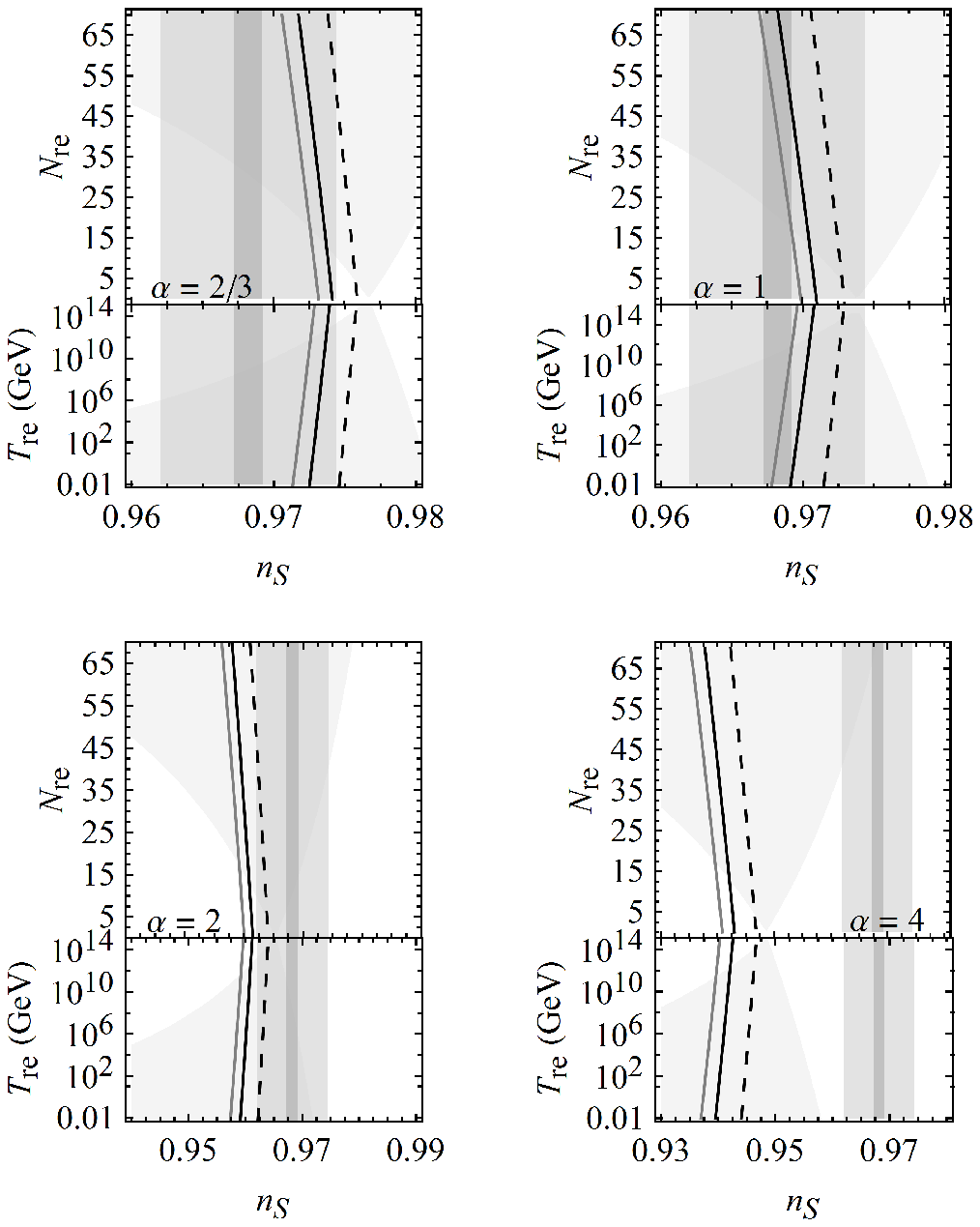}
      \caption{The length of reheating $N_{\re}$ and temperature at the end of reheating $T_\re$, for the model with discontinuous equation of state parameter and polynomial inflationary potential of index $\alpha$. The black dashed, black solid and gray solid curves correspond to $N_\pre$ equals to $10$, $30$ and $40$, respectively and we fixed $\omega_\re=0.25$. The vertical shaded regions are as for figure \ref{fig:Poly_1} and the conic shaded region shows the region delimited in figure \ref{fig:Poly_1} by $-1/3\leq \omega_\re \leq 1$.}
      \label{fig:Poly_2_Npre}
   \end{center}
\end{figure}

\begin{figure}[!ht]
   \begin{center}
      \includegraphics[width=0.75\textwidth]{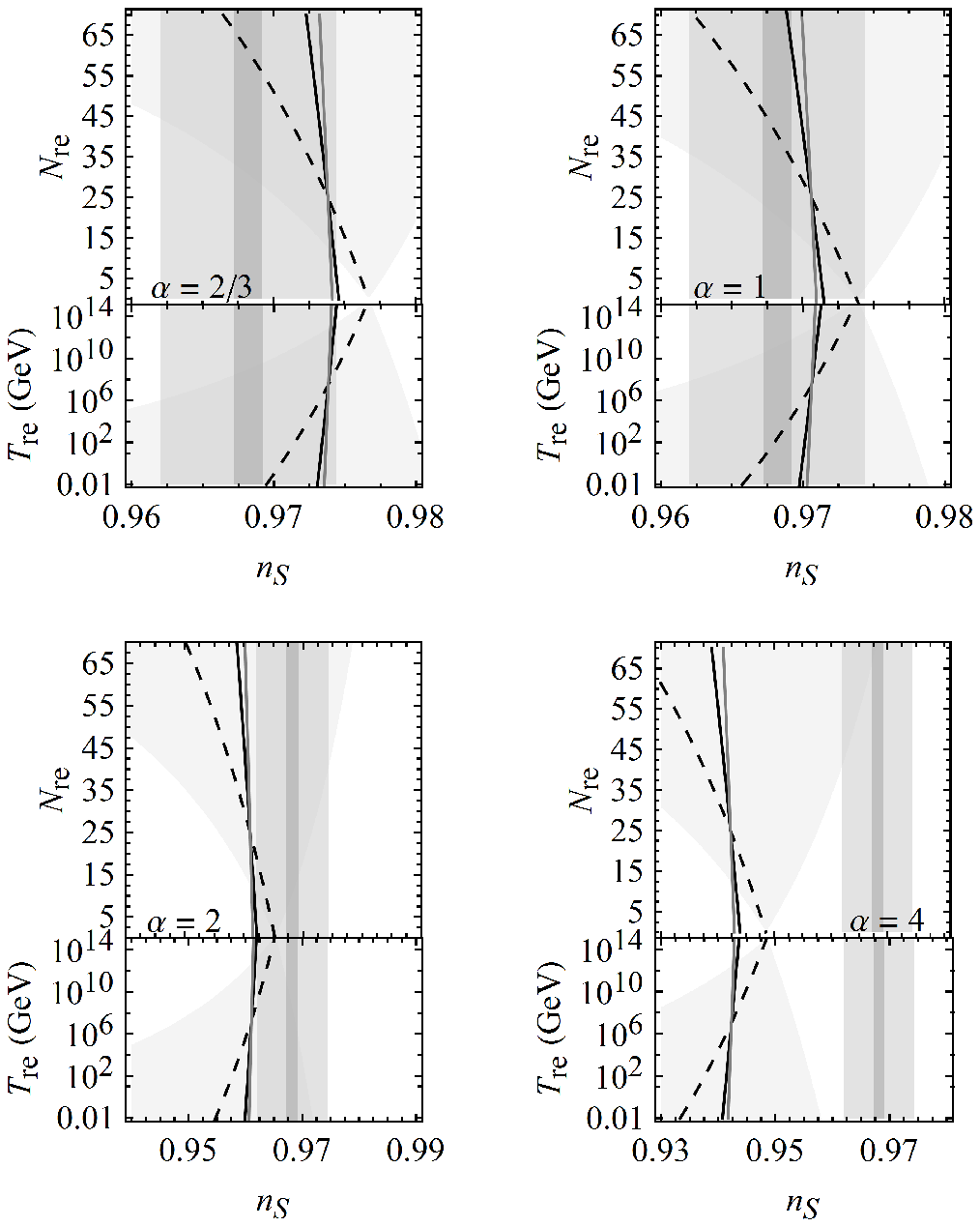}
      \caption{The length of reheating $N_{\re}$ and temperature at the end of reheating $T_\re$, for the model with discontinuous equation of state parameter and polynomial inflationary potential of index $\alpha$. The black dashed, black solid and gray solid curves correspond to $\omega_\re$ equals to $0$, $0.25$ and $0.3$, respectively and we fixed $N_\pre=20$. The vertical shaded regions are as for figure \ref{fig:Poly_1} and the conic shaded region shows the region delimited in figure \ref{fig:Poly_1} by $-1/3\leq \omega_\re \leq 1$.}
      \label{fig:Poly_2_omegare}
   \end{center}
\end{figure}

\begin{figure}[!ht]
   \begin{center}
      \includegraphics[width=0.85\textwidth]{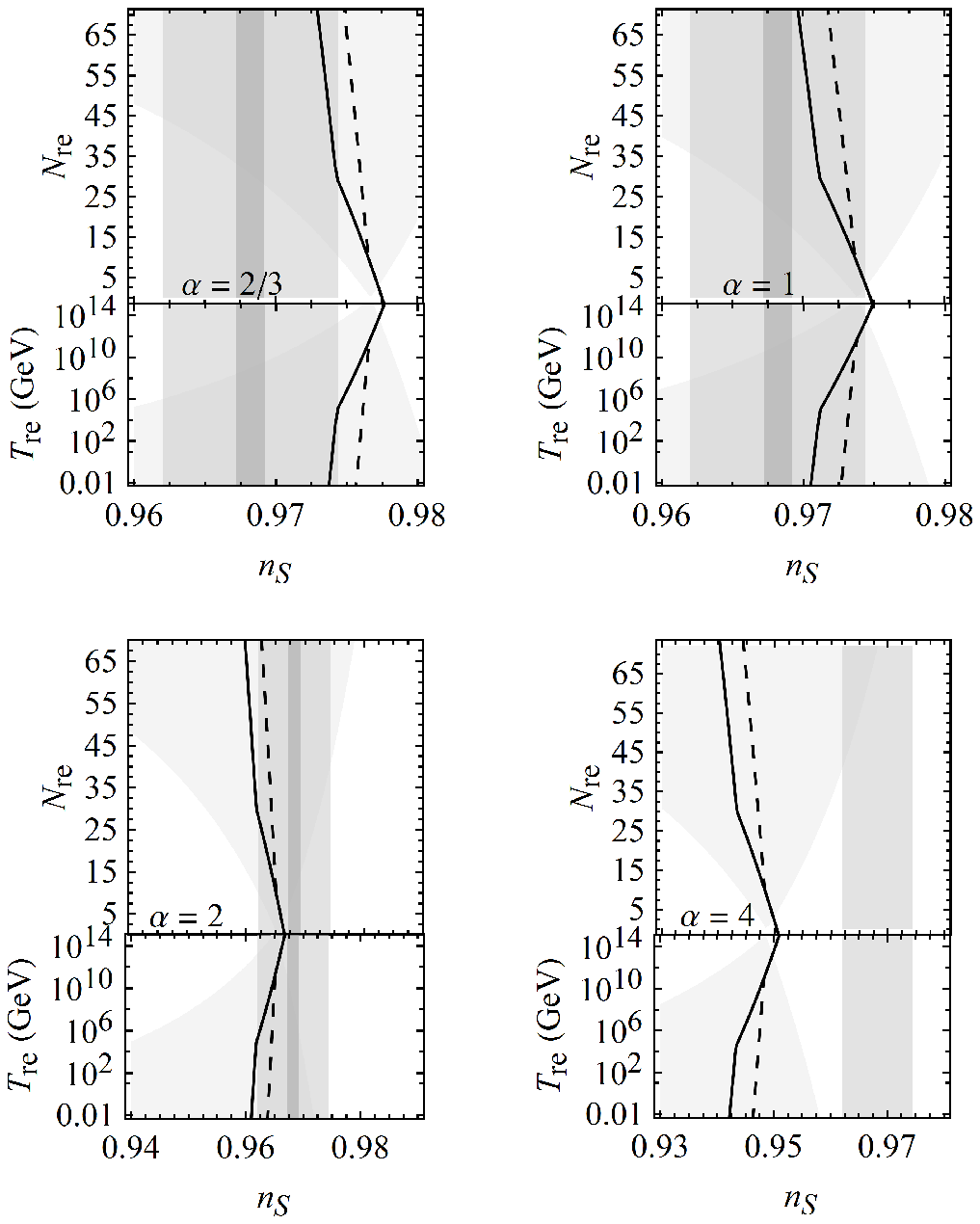}
			\caption{The length of reheating $N_{\re}$ and temperature at the end of reheating $T_\re$, for the time dependent equation of state parameter model and polynomial inflationary potential of index $\alpha$. The black dashed and black solid curves correspond to $N_\pre$ equals to $10$ and $30$, respectively and we fixed $\omega_\re=0.25$. The vertical shaded regions are as for figure \ref{fig:Poly_1} and the conic shaded region shows the region delimited in figure \ref{fig:Poly_1} by $-1/3\leq \omega_\re \leq 1$.}
      \label{fig:Poly_3_Npre}
   \end{center}
\end{figure}

\begin{figure}[!ht]
   \begin{center}
      \includegraphics[width=0.85\textwidth]{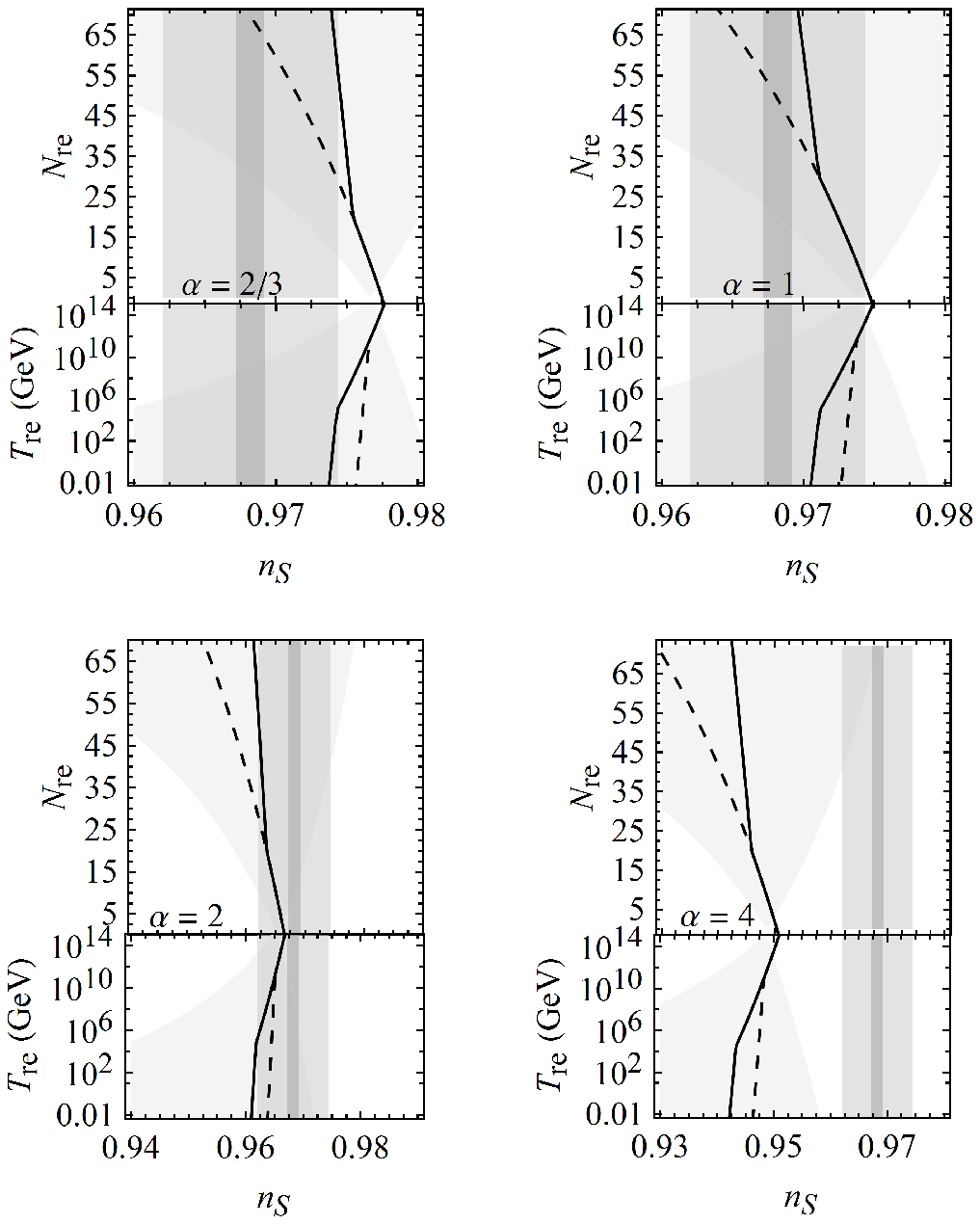}
      \caption{The length of reheating $N_{\re}$ and temperature at the end of reheating $T_\re$, for the time dependent equation of state parameter model and polynomial inflationary potential of index $\alpha$. The black dashed and black solid curves correspond to $\omega_\re$ equals to $0$ and $0.25$, respectively and we fixed $N_\pre=20$. The vertical shaded regions are as for figure \ref{fig:Poly_1} and the conic shaded region shows the region delimited in figure \ref{fig:Poly_1} by $-1/3\leq \omega_\re \leq 1$.}
      \label{fig:Poly_3_omegare}
   \end{center}
\end{figure}

In figures \ref{fig:Poly_1} - \ref{fig:Poly_3_omegare} we plot the reheating length $N_\re$ and the temperature $T_\re$ for $\alpha=2/3$, $1$, $2$ and $4$. From figure \ref{fig:Poly_1}, for the constant equation of state parameter model, we can see that all curves converge to a point where $N_\re \rightarrow 0$, which is defined as the instantaneous reheating limit. This case gives a maximum temperature $T_\re$ at the end of reheating.

Figures \ref{fig:Poly_2_Npre} and \ref{fig:Poly_2_omegare} show $N_\re$ and $T_\re$ for the case of a step function equation of state parameter during reheating, for different values of $N_\pre$ and $\omega_\re$, respectively. In this reheating model the instantaneous reheating point ($N_\re \rightarrow 0$) varies with the free parameters $N_\pre$ and $\omega_\re$.

For the time dependent equation of state parameter case, showed in figures \ref{fig:Poly_3_Npre} and \ref{fig:Poly_3_omegare} for distinct $N_\pre$ and $\omega_\re$, the instantaneous reheating single point behavior is recovered, although it is slightly deviated from the point found in the constant equation of state case.

\subsection{Starobinsky model}
\label{subsec:starobinsky}    

In the case of the Starobinsky inflationary model we begin with the action
\begin{equation}
   S= \int{d^4x\sqrt{-g}\left[\frac{M^2_{\mathrm{Pl}}}{2}\left(R+\alpha R^2\right)\right]} \,,
\end{equation}
where $R$ is the Ricci scalar, $\alpha$ is a model parameter and we considered that other matter fields are sub-dominant during inflation. We use the conformal metric
\begin{equation}
   \bar{g}_{\mu\nu} = \Omega^2g_{\mu\nu} \,,
\end{equation}
where $\Omega^2 \equiv 1 + 2\alpha R$, to perform a conformal transformation that allow us to rewrite the action
\begin{equation}
   S = \int{d^4x\sqrt{-\bar{g}}\left[\frac{M^2_{\mathrm{Pl}}}{2}\bar{R}-\frac{1}{2}(\bar{\partial}\bar{\phi})^2-V(\bar{\phi})\right]} \,,
\end{equation}
with
\begin{equation}
   \label{eq:V_Staro}
   V(\bar{\phi}) = \frac{M^2_{\mathrm{Pl}}}{8\alpha}\left(1-e^{-\sqrt{\frac{2}{3}}\frac{\bar{\phi}}{M_{\mathrm{Pl}}}}\right) \,,
\end{equation}
where we made the definition
\begin{equation}
   \bar{\phi} \equiv \sqrt{\frac{3}{2}}M_{\mathrm{Pl}}\ln{\Omega^2} = \sqrt{\frac{3}{2}}M_{\mathrm{Pl}}\ln{\left(1+2\alpha R\right)} \,.
\end{equation}

From this point we can drop the bar on $\phi$ and with equations (\ref{eq:V_Staro}) and (\ref{eq:Nk}) we can find the number of $e$-folds between the horizon crossing and the end of inflation
\begin{equation}
   \label{eq:Nk_Staro1}
	  N_k \approx \frac{3}{4}e^{\sqrt{\frac{2}{3}}\frac{\phi_k}{M_{\mathrm{Pl}}}} \,,
\end{equation}
where we considered $\phi_k \gg \phi_{\mathrm{end}}$ and $M_{\mathrm{Pl}}e^{\sqrt{\frac{2}{3}}\frac{\phi_k}{M_{\mathrm{Pl}}}} \gg \phi_k$.

Using equations (\ref{eq:epsilon}) and (\ref{eq:eta}) to find the slow-roll parameters $\epsilon$ and $\eta$ as a function of $\phi_k$, we can invert equation (\ref{eq:Nk_Staro1}) to find
\begin{equation}
   \epsilon_k \approx \frac{3}{4N_k^2} \,, \quad \eta_k \approx - \frac{1}{N_k} \,.
\end{equation}
Substituting the above result into equation (\ref{eq:nS}) for $n_S$ and considering $N_k \gg 1$, we will have
\begin{equation}
   \label{eq:Nk_Staro}
	 N_k \approx \frac{2}{1-n_S} \,.
\end{equation}

With equation (\ref{eq:Hk}) we can find
\begin{equation}
   \label{eq:Hk_Staro}
	 H_k \approx \pi M_{\mathrm{Pl}}\sqrt{\frac{3}{2}A_S}(1-n_S) \,.
\end{equation}

Taking into account that at the end of inflation $\epsilon_{\mathrm{end}} = 1$, we write
\begin{equation}
   \label{eq:Vend_Staro}
	 V_{\mathrm{end}} \approx 18\pi^2M^4_{\mathrm{Pl}}A_S\frac{\left(1-n_S\right)^2}{(2+\sqrt{3})^2\left(1-\frac{3}{8}(1-n_S)\right)^2} \,.
\end{equation}

\begin{figure}[!ht]
   \begin{center}
      \includegraphics[width=0.7\textwidth]{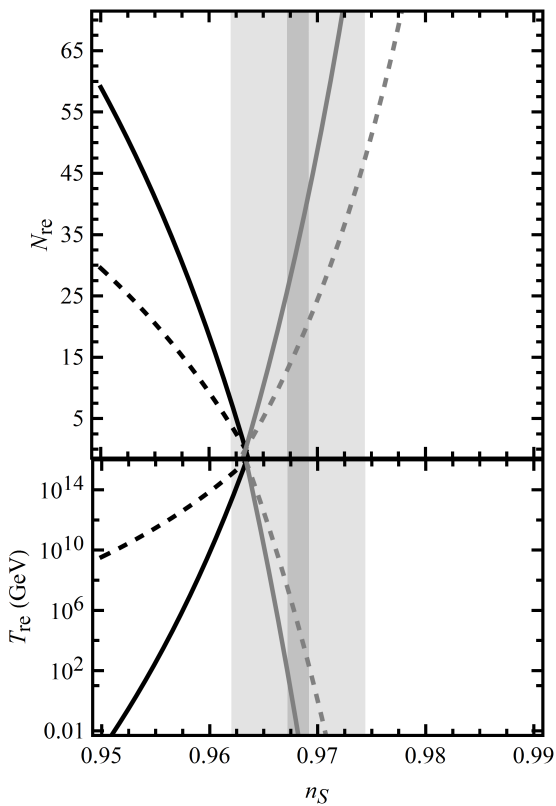}
		  \caption{The length of reheating $N_{\re}$ and temperature at the end of reheating $T_\re$, for the model with constant equation of state parameter and both Starobinsky and Higgs inflation. The black dashed, black solid, gray dashed and gray solid curves correspond to $\omega_\re$ equals to $-1/3$, $0$, $2/3$ and $1$, respectively. The light gray shaded region represents the $1\sigma$ bounds on the primordial power spectrum index $n_{\mathrm S}$ from Planck, while the dark gray shaded region correspond to the $1\sigma$ bound of a CMB experiment with sensitivity $\pm 10^{-3}$ and the same central value for $n_{\mathrm S}$ as Planck.}
      \label{fig:Staro_1}
   \end{center}
\end{figure}

\begin{figure}[!ht]
   \begin{center}
      \includegraphics[width=0.7\textwidth]{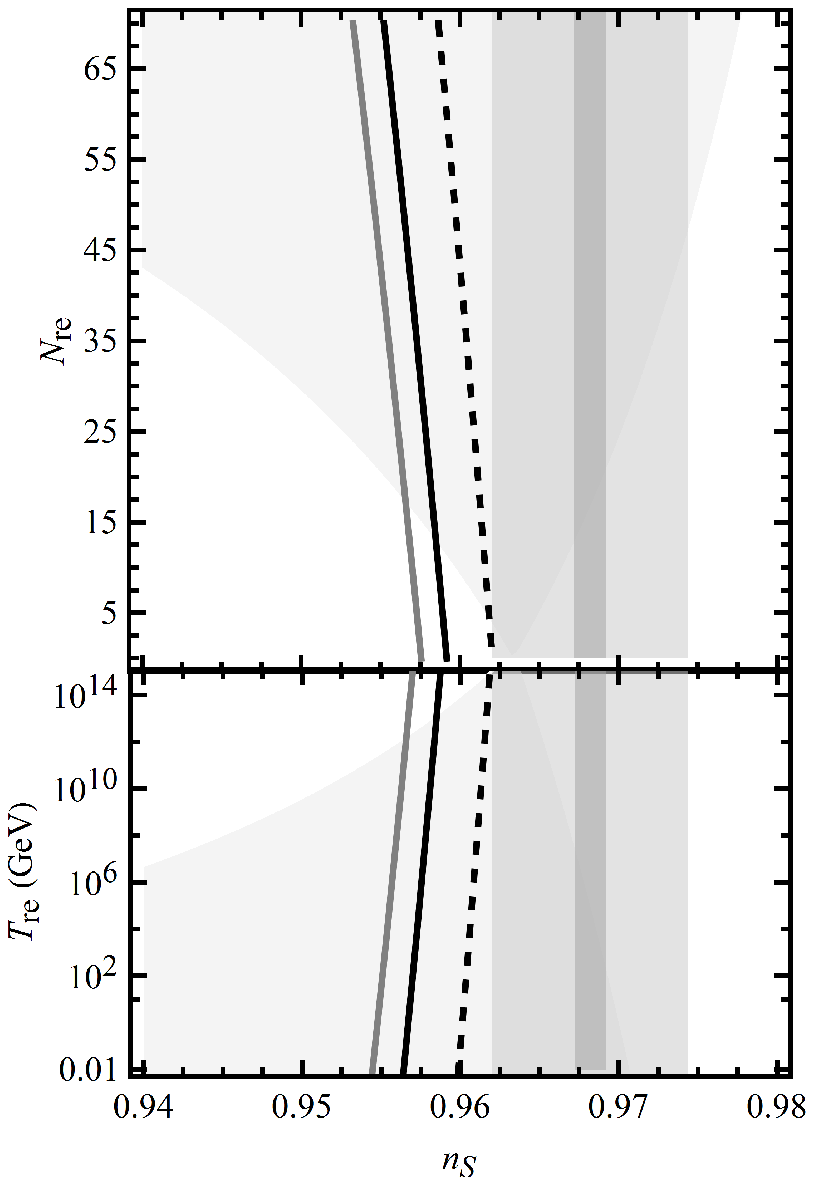}
      \caption{The length of reheating $N_{\re}$ and temperature at the end of reheating $T_\re$, for the model with discontinuous equation of state parameter and both Starobinsky and Higgs inflation. The black dashed, black solid and gray solid curves correspond to $N_\pre$ equals to $10$, $30$ and $40$, respectively and we fixed $\omega_\re=0.25$. The vertical shaded regions are as for figure \ref{fig:Staro_1} and the conic shaded region shows the region delimited in figure \ref{fig:Staro_1} by $-1/3\leq \omega_\re \leq 1$.}
      \label{fig:Staro_2_Npre}
   \end{center}
\end{figure}

\begin{figure}[!ht]
   \begin{center}
      \includegraphics[width=0.7\textwidth]{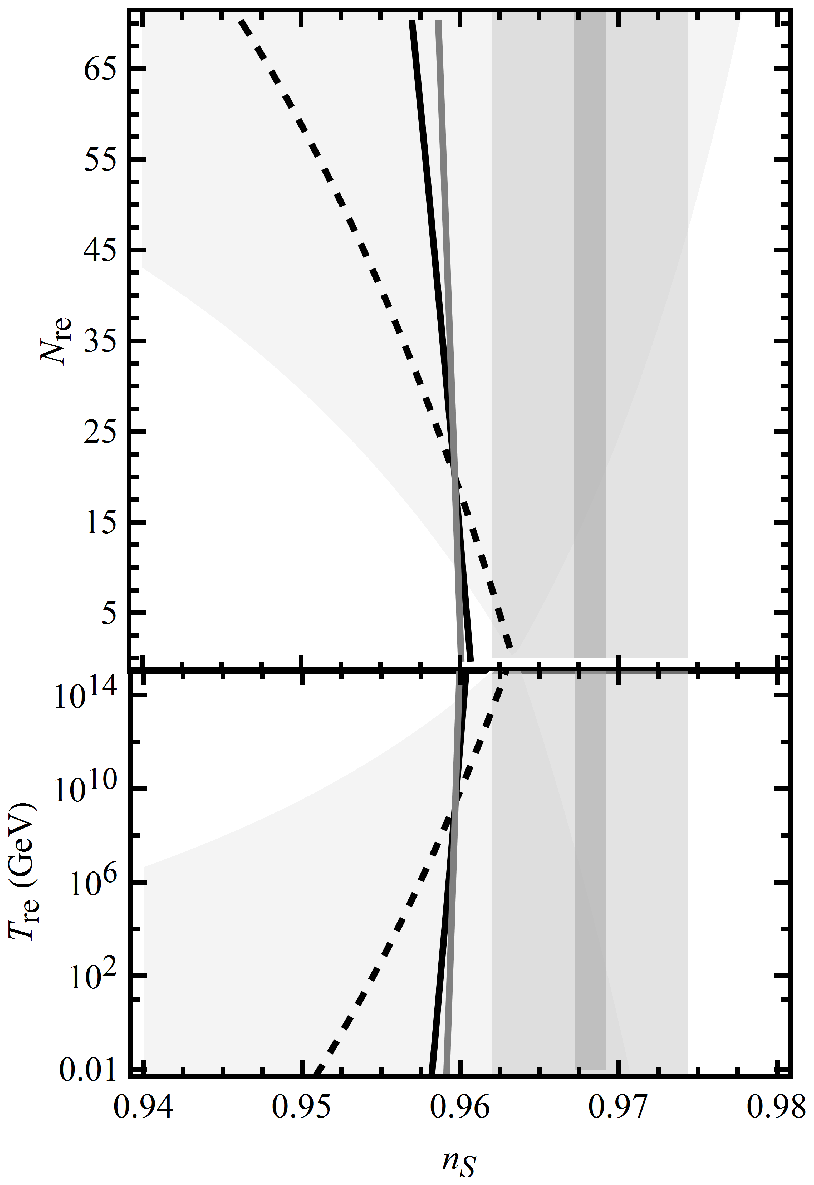}
      \caption{The length of reheating $N_{\re}$ and temperature at the end of reheating $T_\re$, for the model with discontinuous equation of state parameter and both Starobinsky and Higgs inflation. The black dashed, black solid and gray solid curves correspond to $\omega_\re$ equals to $0$, $0.25$ and $0.3$, respectively and we fixed $N_\pre=20$. The vertical shaded regions are as for figure \ref{fig:Staro_1} and the conic shaded region shows the region delimited in figure \ref{fig:Staro_1} by $-1/3\leq \omega_\re \leq 1$.}
      \label{fig:Staro_Nre2_omegare}
   \end{center}
\end{figure}

\begin{figure}[!ht]
   \begin{center}
      \includegraphics[width=0.7\textwidth]{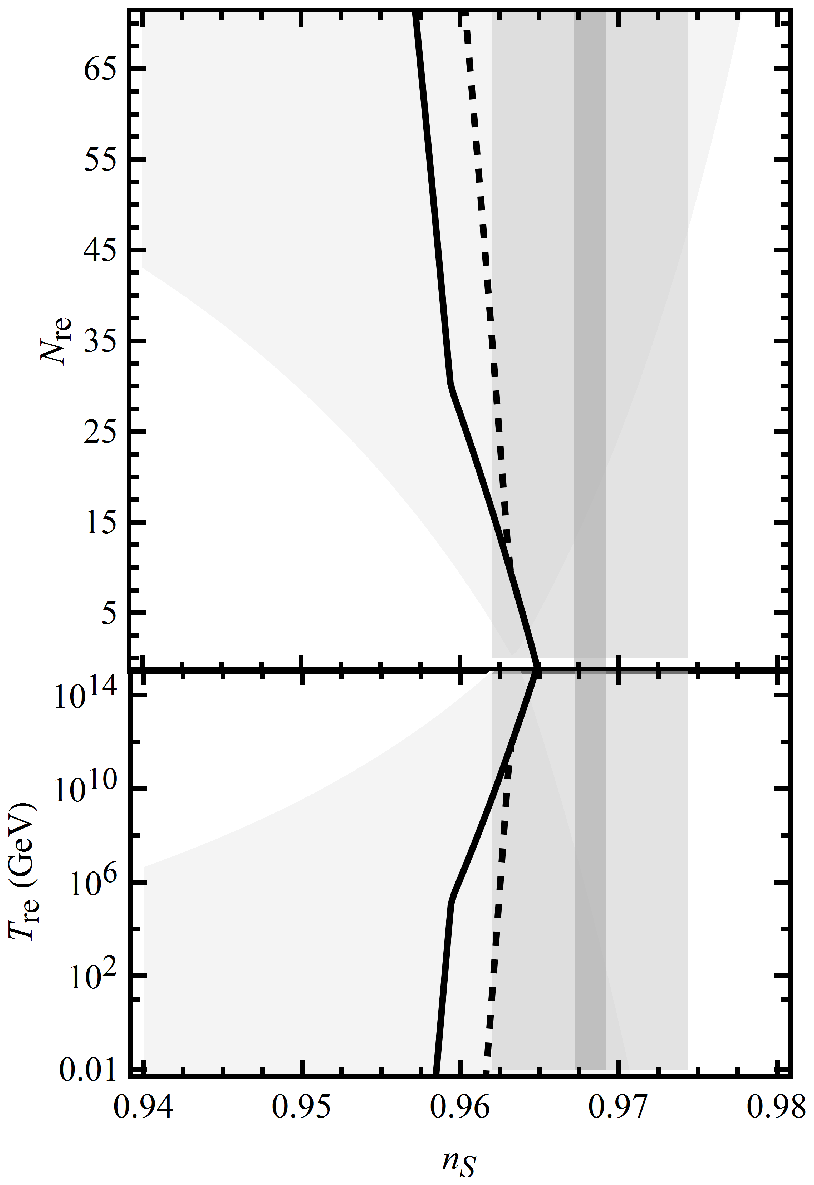}
			\caption{The length of reheating $N_{\re}$ and temperature at the end of reheating $T_\re$, for the time dependent equation of state parameter model and both Starobinsky and Higgs inflation. The black dashed and black solid curves correspond to $N_\pre$ equals to $10$ and $30$, respectively and we fixed $\omega_\re=0.25$. The vertical shaded regions are as for figure \ref{fig:Staro_1} and the conic shaded region shows the region delimited in figure \ref{fig:Staro_1} by $-1/3\leq \omega_\re \leq 1$.}
      \label{fig:Staro_3_Npre}
   \end{center}
\end{figure}

\begin{figure}[!ht]
   \begin{center}
      \includegraphics[width=0.7\textwidth]{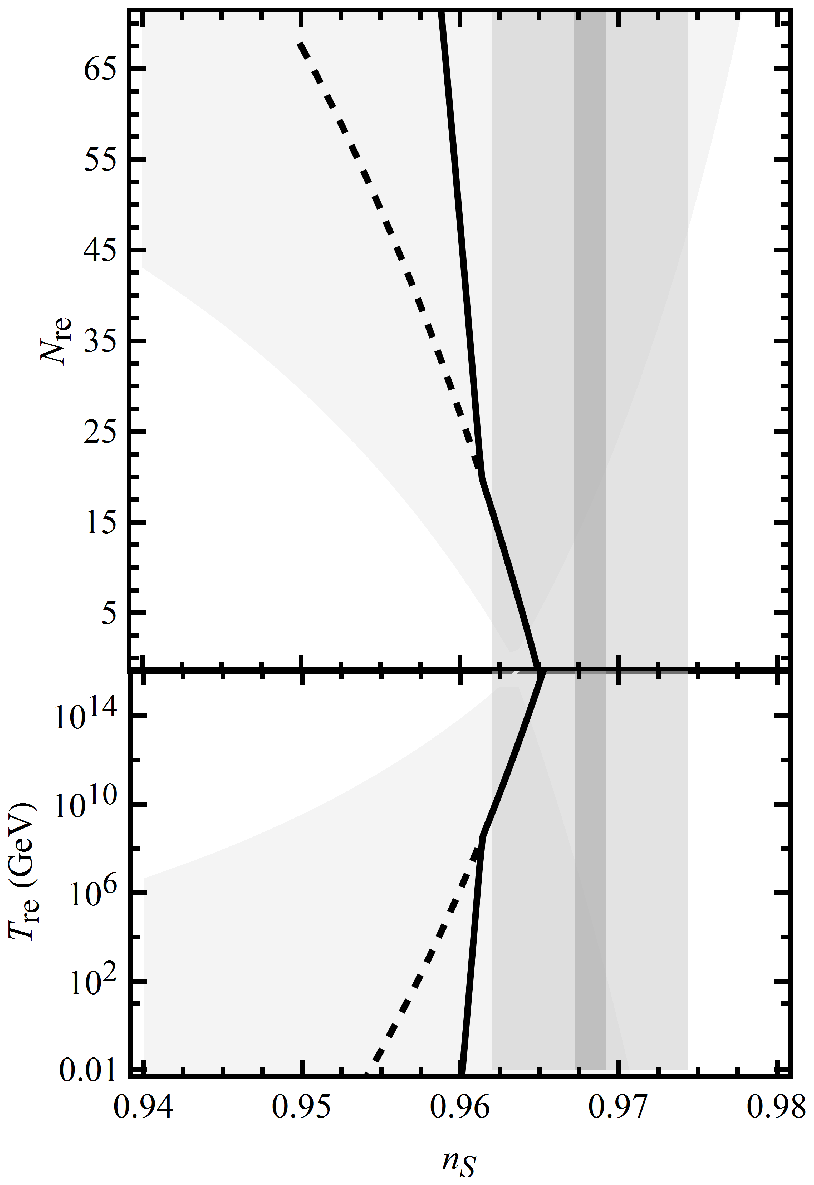}
      \caption{The length of reheating $N_{\re}$ and temperature at the end of reheating $T_\re$, for the time dependent equation of state parameter model and both Starobinsky and Higgs inflation. The black dashed and black solid curves correspond to $\omega_\re$ equals to $0$ and $0.25$, respectively and we fixed $N_\pre=20$. The vertical shaded regions are as for figure \ref{fig:Staro_1} and the conic shaded region shows the region delimited in figure \ref{fig:Staro_1} by $-1/3\leq \omega_\re \leq 1$.}
      \label{fig:Staro_3_omegare}
   \end{center}
\end{figure}

The result of plugging equations (\ref{eq:Nk_Staro})--(\ref{eq:Vend_Staro}) into the models described by the equations (\ref{eq:Nre1}), (\ref{eq:Nre2}) and (\ref{eq:Nre3}) can be seen in figures \ref{fig:Staro_1}--\ref{fig:Staro_3_omegare}.

Figures \ref{fig:Staro_1} and \ref{fig:Staro_3_Npre}--\ref{fig:Staro_3_omegare} show the consistency between the Starobinsky model and Planck's 1$\sigma$ bound for the constant equation of state and the time dependent equation of state models, respectively. The instantaneous reheating behavior of the constant equation of state model is again recovered in the case of the time dependent equation of state.

\subsection{Higgs Inflation}
\label{subsec:higgs}        

In Higgs inflation the Higgs field $h$ is non-minimally coupled to gravity, and has the following action
\begin{equation}
   S = \int{d^4x\sqrt{-g}\left[\frac{M^2_{\mathrm{Pl}}}{2}\left(1+2\xi\frac{h^2}{M^2_{\mathrm{Pl}}}\right)R-\frac{1}{2}\left(\partial h\right)^2 - \frac{\lambda}{4}\left(h^2-v^2\right)^2\right]} \,,
\end{equation}
where we considered other matter fields sub-dominant during inflation.

We make a conformal transformation
\begin{equation}
   \bar{g}_{\mu\nu} = \Omega^2g_{\mu\nu} \,,
\end{equation}
where $\Omega^2 \equiv 1+2\xi\frac{h^2}{M^2_{\mathrm{Pl}}}$, which will lead us to the action
\begin{equation}
   S = \int{d^4x\sqrt{-\bar{g}}\left[\frac{M^2_{\mathrm{Pl}}}{2}\bar{R}-\frac{1}{2}(\bar{\partial}\bar{h})^2-V(\bar{h})\right]} \,,
\end{equation}
with
\begin{equation}
   \label{eq:V_Higgs}
   V(\bar{h}) = \frac{\lambda M^4_{\mathrm{Pl}}}{4\xi^2}\left(1-e^{-\sqrt{\frac{2}{3}}\frac{\bar{h}}{M_{\mathrm{Pl}}}}\right) \,,
\end{equation}
where we use the unitary gauge in order to canonically normalize all the four degrees of freedom of the Higgs field, which yields the following transformation
\begin{equation}
   \bar{h} = \sqrt{\frac{3}{2}}M_{\mathrm{Pl}}\ln{\left(1+\frac{\xi h^2}{M^2_{\mathrm{Pl}}}\right)} \,.
\end{equation}

Making the identification $\alpha = \frac{\xi^2}{2\lambda M^2_{\mathrm{Pl}}}$ in equation (\ref{eq:V_Higgs}) we see that the predictions made here for the Higgs model are the same as for the Starobinsky model, since our equations do not depend on the proportionality constants of the potentials (\ref{eq:V_Higgs}) and (\ref{eq:V_Staro}). These models are equivalent for large values of the fields, but its small-field behavior are different. However, in ref. \cite{riot} the authors show that, in particular, there is a good agreement between the observational data and both Starobinsky and Higgs's inflationary models, both valid up to Planckian scales.

\section{Conclusions}        
\label{sec:conclusions}      

Since the reheating process has an important contribution to establish the thermal history of the Universe after inflation it can indirectly  affect the predictions for inflation. The hypothesis that the equation of state parameter during reheating rises from $0$ to $1/3$ in three different manners gave us reheating constraints on parameters that can be probed using CMB data. By taking into account the expansion history of the Universe between the time when the CMB modes left the horizon and the time of observation we are lead to three different relations between reheating and inflationary parameter from a specific model, determined by the adopted model for equation of state parameter of the reheating stage.

Models with constant equation of state parameter have already been studied in the literature and our novel contribution is the analysis with time varying models. First, we adopt a model where the equation of state parameter changes sharply from $0$ to values close to $1/3$. In the second model this change in $\omega_\re$ is continuously made. In both cases the price to pay is the presence of two knew model parameters. From equations (\ref{eq:Nre1}), (\ref{eq:Nre2}) and (\ref{eq:Nre3_1}) one can clearly see the equivalence between the three models. In equations (\ref{eq:kappa_infity}) and (\ref{eq:Nre3_approx}) we show that it is possible to recover the discontinuous case from the continuous one and, more important, figure \ref{fig:Poly_3_kappa} shows that our final result for the continuous case do not depends on $\kappa$, that is the extra model parameter that controls how sharp is the jump of the equation of state parameter during reheating.

In this work we considered $\omega_\pre = 0$ and $0\leq \omega_\re \leq 1/3$, since we think it is more consistent with the result found in \cite{Podolsky:2005bw}. The main effect of this choice is that the regions between our curves in the $N_\re ~\times~ n_S$ and $T_\re ~\times~ n_S$ planes are much more narrow when compared with the ones that are found in the literature, which consider a constant equation of state parameter in the range $-1/3\leq \omega_\re \leq 1$. For the discontinuous equation of state parameter case the curves do not converge to the same point in case of instantaneous reheating ($N_\re \rightarrow 0$), but this feature can be recovered for the continuous jump, although this point is very slightly dislocated in favor of a bigger scalar spectral index $n_S$. In both continuous and discontinuous jumps there is a bend in the trajectory of the curves, which coincides with the fixed value of $N_\pre$, that is the point where the equation of state parameter changes its value from $\omega_\pre = 0$ to $\omega_\re=0.2\textup{ -- }0.3$.

The present constraints for the inflationary parameters are consistent with various models for inflation and allows a broad range of reheating temperatures. Although we showed that, as claimed in the literature, to consider a constant equation of state parameter during reheating can be a good approximation, we also saw that taking into account a varying equation of state can bring some small features to the model.
\vspace{1cm}

\textbf{Acknowledgments:}
S.~V.~B.~Gon\c{c}alves thanks CNPq (Brazil) and both authors thank FAPES (Brazil) for the partial financial support.

\end{document}